\documentclass[prb,a4paper, secnumarabic, twocolumn, amssymb, nobibnotes, aps]{revtex4-2}

\usepackage{graphicx}
\usepackage{pslatex}
\usepackage{xspace}
\usepackage{amsmath}
\usepackage{color}
\usepackage{booktabs}

\newcommand{\etal}{\emph{et\,al.}\xspace}

\begin{document}

\title{Nongeminate and Geminate Recombination in PTB7:PC$_{71}$BM solar cells}

\author{A.~Foertig$^1$}
\author{J.~Kniepert$^{2}$}
\author{M.~Gluecker$^1$}
\author{T.~Brenner$^2$}
\author{V.~Dyakonov$^{1,3}$}\email{dyakonov@physik.uni-wuerzburg.de}
\author{D.~Neher$^{2}$}
\author{C.~Deibel$^1$}\email{deibel@disorderedmatter.eu}

\affiliation{$^1$ Experimental Physics VI, Julius-Maximilians-University of W\"urzburg, D-97074 W\"urzburg, Germany}
\affiliation{$^2$ Institute of Physics and Astronomy, University of Potsdam, Karl-Liebknecht-Str. 24-25, 14476 Potsdam, Germany}
\affiliation{$^3$ Bavarian Center for Applied Energy Research e.V. (ZAE Bayern), D-97074 W\"urzburg, Germany}

\date{August 11, 2013}

\begin{abstract}
A combination of transient photovoltage (TPV), voltage dependent charge extraction (CE) and time delayed collection field (TDCF) measurements is applied to poly[[4,8-bis[(2-ethylhexyl)oxy]benzo[1,2-b:4,5-b']dithiophene-2,6-diyl] [3-fluoro-2-[(2-ethylhexyl)carbonyl] thieno[3,4-b]thiophenediyl]] (PTB7):[6,6]-phenyl-C71-butyric acid (PC$_{71}$BM) bulk heterojunction solar cells to analyze the limitations of photovoltaic performance. Devices are processed from pure chlorobenzene (CB) solution and a subset was optimized with 1,8-diiodooctane (DIO) as co-solvent. The dramatic changes in device performance are discussed with respect to the dominating loss processes. While in the devices processed from CB solution, severe geminate and nongeminate recombination is observed, the use of DIO facilitates efficient polaron pair dissociation and minimizes geminate recombination. Thus, from the determined charge carrier decay rate under open circuit conditions and the voltage dependent charge carrier densities $n(V)$, the nongeminate loss current $j_{loss}$ of the samples with DIO alone enables us to reconstruct the current/voltage ($j/V$) characteristics across the whole operational voltage range. Geminate and nongeminate losses are considered to describe the $j/V$ response of cells prepared without additive, but lead to a clearly overestimated device performance. We attribute the deviation between measured and reconstructed $j/V$ characteristics to trapped charges in isolated domains of pure fullerene phases. 

\textbf{This is the pre-peer reviewed version of the following article:\\
A.~Foertig, J.~Kniepert, M.~Gluecker, T.~Brenner, V.~Dyakonov, D.~Neher, and C.~Deibel.
Nongeminate and geminate recombination in PTB7:PC71BM solar cells.
Adv. Funct. Mater.~24:1306 (2014). DOI:~10.1002/adfm.201302134}
\end{abstract}


\maketitle

\section{Introduction}
In recent years the power conversion efficiency (PCE) of organic solar cells (OSC) based on polymer:fullerene mixtures improved above 11\%~\cite{green2013, li2012}. This performance enhancement is mainly due to the development of new low bandgap semiconductors and their broadened absorption spectrum. Thus, these new material compositions deserve particular consideration to advance the understanding of the crucial steps from photon absorption to photocurrent. In order to further improve device performance, identifying the performance limiting loss mechanisms is essential. On the one hand, geminate losses of bound electron--hole pairs compete with polaron pair dissociation. As electron--hole dissociation via an intermediate charge transfer state might require a certain activation energy to obtain free charges (see Ref.~\onlinecite{bredas2009}) this process can be supported by an external electric field. On the other hand, nongeminate recombination of free--free or trapped--free polarons after successful polaron pair dissociation depends on the charge accumulation in the device and, thus, relies on the applied voltage and respective current flow.
It was recently shown for polymer and small molecule based OSC that both, nongeminate~\cite{shuttle2010, gluecker2012} as well as geminate losses,~\cite{credgington2012, albrecht2012b} can have a strong impact on the device performance and the shape of the $j/V$ characteristics, depending on the photoactive material.

In the present study, we analyzed organic solar cells based on PTB7:PC$_{71}$BM blends processed from a solution of pure chlorobenzene and from a combination of chlorobenzene and the co-solvent 1,8-diiodooctane (DIO). This additive selectively dissolves the fullerene~\cite{lou2011} and leads to a dramatic improvement of device performance and to a change in recombination dynamics. We find that while the devices processed from CB with additive are purely dominated by nongeminate recombination, the ones processed from CB alone suffer from both, geminate and nongeminate recombination pathways. For samples processed from pure CB, $j/V$ reconstruction from recombination dynamics---an approach recently introduced in Ref.~\onlinecite{shuttle2010} for organic solar cells and applied in Ref.~\onlinecite{gluecker2012}---deviates from the directly measured $j/V$ characteristics. We propose that a considerable amount of trapped charges $n_t$ in isolated fullerene domains---exceeding the density of free charges $n_c$---is responsible for this discrepancy. We discuss our results in terms of the change of active layer morphology upon the use of an additive~\cite{collins2012, kraus2012}.

\section{Results}
\begin{figure}[tb]
	\includegraphics[width=0.9\linewidth]{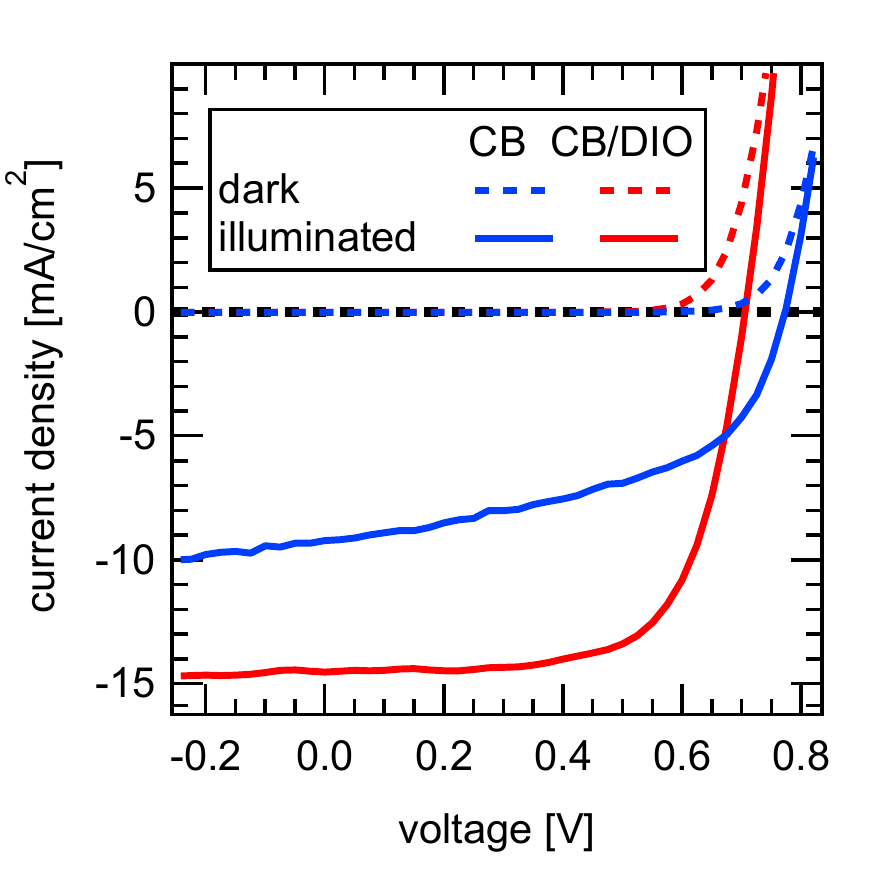}
	\caption{Measured dark (dashed) and illuminated (solid) current/voltage response of PTB7:PC$_{71}$BM solar cells processed from chlorobenzene solution and from a combination of chlorobenzene and DIO. Illumination intensity was set to 1 sun under a simulated AM1.5G spectrum and the temperature was 300K.}
	\label{fig:iv}
\end{figure}

Two sets of organic solar cells based on a PTB7:PC$_{71}$BM 1:1.5 blend were fabricated differing only by the use of a solvent additive (see experimental section). The influence of the co-solvent DIO on device performance becomes evident in the respective current/voltage response depicted in Fig.~\ref{fig:iv}. Significant improvement of fill factor and photocurrent is observed when using the additive, in accordance with previous findings~\cite{liang2010, brenner2011, hammond2011, collins2012}. Under one sun simulated illumination, the sample processed without additive yields an open circuit voltage of $V_{oc}=770$~mV, a short circuit current density of $j_{sc}=9.2$~mA/cm$^2$ with a fill factor of $\emph{FF}=51$\% and a device efficiency of $\eta=3.6$\%. In contrast, the device processed with DIO shows $V_{oc}=710$~mV, $j_{sc}=14.6$~mA/cm$^2$, a fill factor of $\emph{FF}=67$\% and an almost twice as high PCE of $\eta=7.0$\%. The chosen additive DIO is rather effective to influence the PTB7:PC$_{71}$BM composite, as it has a high boiling point and selectively dissolves PC$_{71}$BM, which was reported in Ref.~\onlinecite{lou2011} to enable fullerene intercalation into the polymer network during film formation.

In the following, the differences in nongeminate and geminate recombination dynamics for both device types are presented in order to investigate the origin of the dramatic change of $j_{sc}$ and \emph{FF}.

\begin{figure}[tb]
	\includegraphics[width=0.9\linewidth]{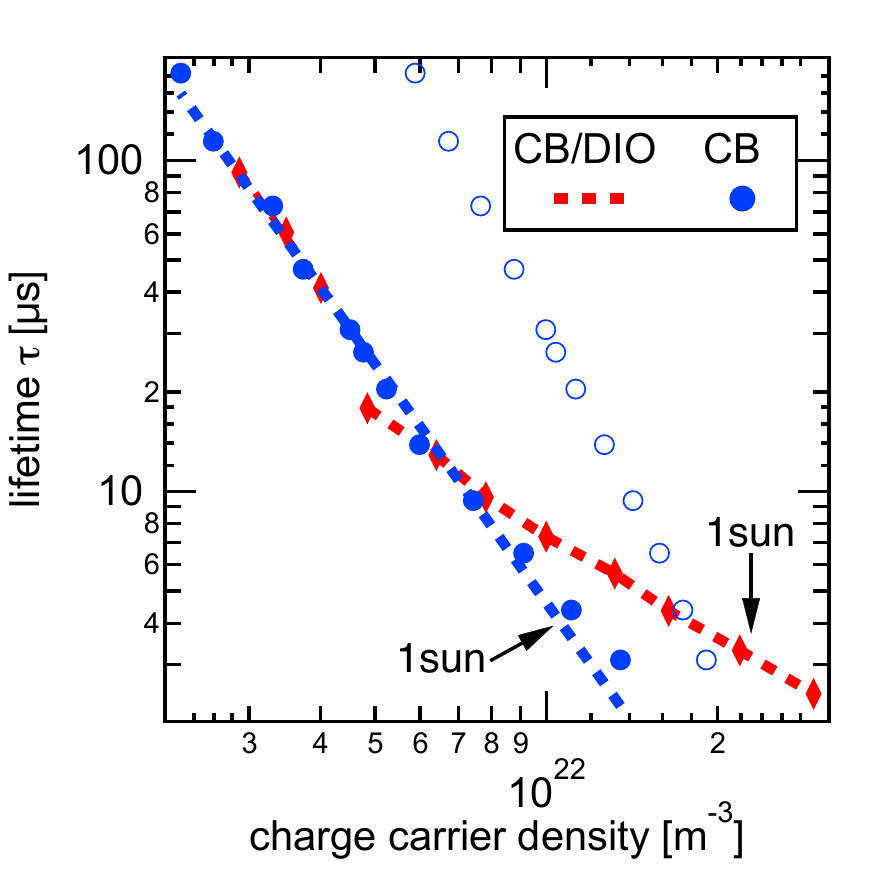}
	\caption{Charge carrier lifetime $\tau$ derived from TPV measurements as a function of charge carrier density $n$ obtained from CE for PTB7:PC$_{71}$BM solar cells processed from chlorobenzene solution and from chlorobenzene and DIO. Diamonds and filled circles represent experimental data points, while the dashed line demonstrates the fit according to $\tau(n)=\tau_0n^{-\lambda}$. The data points represented by blue open circles are based on a charge carrier density reconstruction described later in the text.}
	\label{fig:tau_n}
\end{figure}

The nongeminate recombination rate $R$ of photogenerated charge carriers can empirically be defined as~\label{page:R-at-Voc}
\begin{equation}
R(n)=\frac{n}{\tau(n)}\propto n^{(\lambda+1)},
\label{eq:nong}
\end{equation}
with the charge carrier density $n$, the effective charge carrier lifetime $\tau(n$), and $\lambda +1$ representing the apparent nongeminate recombination order. In order to determine the nongeminate recombination rate experimentally, transient photovoltage (TPV) and charge extraction (CE) experiments under open circuit ($V_{oc}$) conditions were applied. TPV is based on monitoring the photovoltage decay upon a small optical perturbation during various constant bias light conditions~\cite{shuttle2008}. The small perturbation charge carrier lifetime $\tau_{\Delta n}$ is extracted from the exponential voltage decay in dependence of the respective open circuit voltage at various illumination levels. As shown earlier, the parameter $\tau_{\Delta n}$ is related to the total charge carrier lifetime by $\tau(n)=\tau_{\Delta n}(\lambda+1$) (see Eq.~(\ref{eq:nong}))~\cite{hamilton2010, gluecker2012}. The average charge carrier density $n$ under open circuit conditions is accessible from CE ($n(V_{oc})$). The data are corrected for the geometric capacitance  (see further details in discussion) and, iteratively, for charges lost due to recombination during extraction~\cite{shuttle2008b,gluecker2012}. In Fig.~\ref{fig:tau_n}, TPV and CE are combined to yield the charge carrier lifetime $\tau$ as a function of charge carrier density under open circuit conditions, $\tau(n)$. 

For both device types, a power law dependence according to $\tau(n)=\tau_0n^{-\lambda}$ is found: In the low charge carrier density regime, the $\tau(n)$ values coincide, yielding the same slope ($\lambda\approx 2.4$) for both devices. However, at higher carrier densities the device with additive shows a reduced slope with $\lambda\approx1.1$. For the device without DIO only a minor change in slope becomes apparent. According to Ref.~\onlinecite{kirchartz2011}, an activation energy $E_u$ characterizing the exponential trap distribution can be estimated from the parameter $\lambda$~\cite{foertig2012b}. For low charge carrier densities we find $E_u\approx 2\lambda \cdot kT/n_{\tau}=51$~meV. This indicates trap limited recombination for the device processed without additive and in the low charge carrier regime of the device prepared from CB/DIO. Above $n=10^{22}~m^{-3}$ direct (Langevin--type) recombination of free charge carriers is expected, as the apparent recombination order $\lambda+1$ is about 2 at 300K. For an illumination equivalent to 1 sun, the nongeminate charge carrier lifetime for the device processed without additive is slightly longer than for the one with additive (see arrows in Fig.~\ref{fig:tau_n}).

\begin{figure}[tb]
	\includegraphics[width=0.9\linewidth]{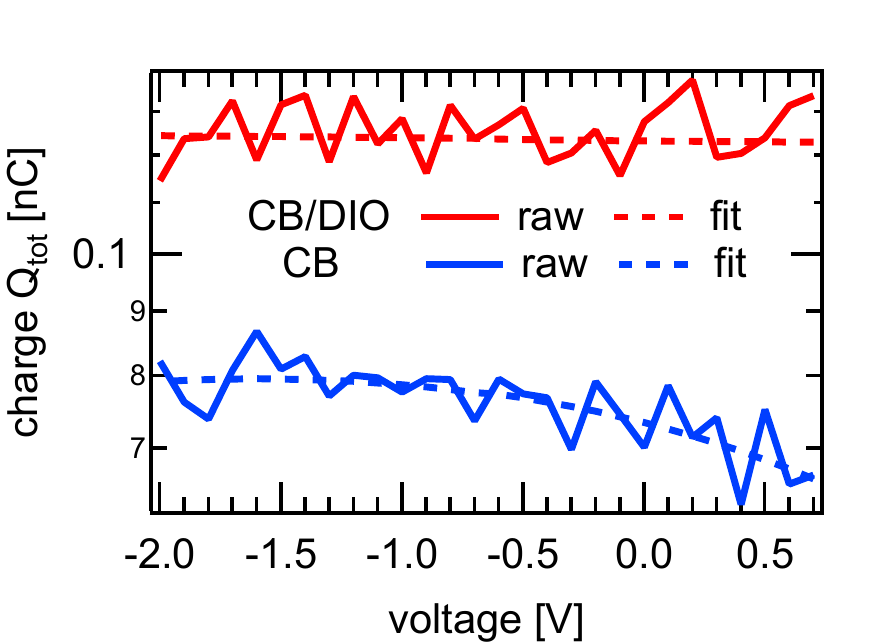}
	\caption{Measured raw data and fit (dashed lines) of total extracted charge $Q_{tot}$ from TDCF measurements in dependence on the applied prebias voltage.}
	\label{fig:tdcf}
\end{figure}

Time delayed collection field (TDCF) measurements are a proven method to study charge photogeneration~\cite{albrecht2012b, mingebach2012}. Thus, by investigating the total extracted charge $Q_{tot}$ under a range of prebias voltages the influence of geminate recombination is analyzed.  These experiments are performed under very low fluences of 0.05 $\mu$J/cm$^2$ and a short delay of 10~ns to minimize nongeminate recombination during charge carrier extraction~\cite{kniepert2011}. Free carrier generation in competition to geminate recombination is completed within a few nanoseconds as found by transient absorption experiments~\cite{howard2010}. In Fig.~\ref{fig:tdcf} the total charge $Q_{tot}$ is plotted versus the prebias voltage applied during the delay time $t_d$. Thereby, a polynomial fit of 3rd order over the complete voltage range was used to smooth the experimental raw data. While the extracted total charge is rather constant for the device with additive, a clear voltage dependence for the sample processed from CB becomes apparent. In the latter case, $Q_{tot}$ decreases by nearly 20\% when going from reverse bias to $V_{oc}$. A voltage dependent photogeneration is caused by geminate recombination competing with field-assisted free charge generation, which is quantified by the polaron pair dissociation probability PP($V$). The above described polynomial fit was used to find an analytical approximation for PP($V$) between short and open circuit conditions. The data in Fig.~\ref{fig:tdcf} also shows that more charge is generated in the DIO-processed blend throughout the entire bias range. This finding is consistent with the larger $j/V$ response of this blend under illumination.

\begin{figure}[tb]
	\includegraphics[width=0.9\linewidth]{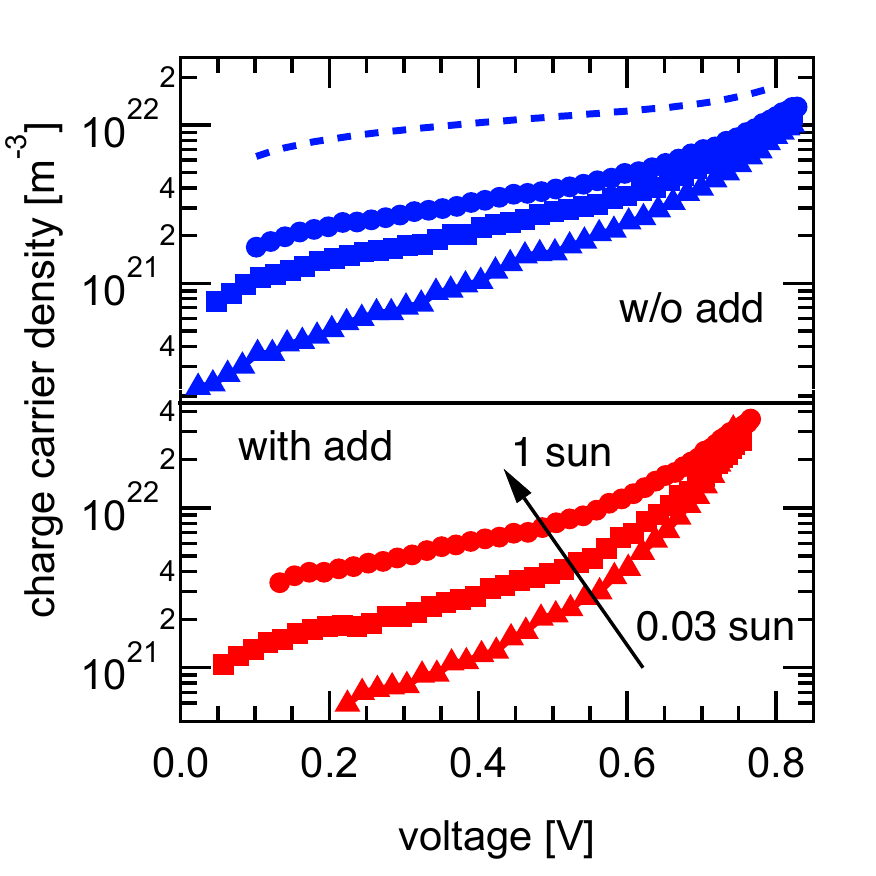}
	\caption{Voltage dependent charge carrier density $n(V)$ from charge extraction experiments for PTB7:PC$_{71}$BM devices with and without additive at three different light intensities.}
	\label{fig:n_V}
\end{figure}

In order to understand the impact of geminate and nongeminate recombination on the performance of the device, a procedure known as $j/V$ reconstruction was applied, in analogy to recent studies on P3HT:PC$_{61}$BM~\cite{shuttle2010, gluecker2012}. In order to describe the steady state current density $j/V$ of the device, the continuity equation for charge carriers
\begin{equation}
	\frac{1}{q}\left(\frac{\emph{dj}}{\emph{dx}}\right)+G-R=0
	\label{eq:con}
\end{equation}
was integrated, assuming spatially constant rates for generation ($G$) and recombination ($R$). This yields
\begin{equation}
	j(V)=qdG-qdR=j_{gen}(V)-j_{loss}(V),
	\label{eq:j}
\end{equation}
with the elementary charge $q$, the device thickness $d$, the respective generation current $j_{gen}$ and nongeminate loss current $j_{loss}$. In order to calculate the voltage dependence $j_{loss}$($V$) from $V=0$~V to $V_{oc}$, CE experiments under the desired voltage were performed, in analogy to the measurements at $V_{oc}$ described above.
We point out that all voltages were corrected for the series resistance $R_s$ by calculating $V=V_{app}-R_sI$. From the ohmic range of the dark $j/V$ curve, the values $R_s\approx 84~\Omega$ for the device with additive and $R_s=105~\Omega$ for the one without additive were derived. The voltage dependent charge carrier density for both devices is shown in Fig.~\ref{fig:n_V} for three different light intensities.

The $n(V)$ relation and the dependence of $\tau$ on $n$ found under $V_{oc}$ conditions, Fig.~\ref{fig:tau_n}, are used to calculate the charge carrier density dependent recombination rate $R(n(V))$ for the respective voltage by Eq.~(\ref{eq:nong}). This data were fed into Eq.~(\ref{eq:j}), which allowed to determine the nongeminate recombination current $j_{loss}(n(V))$. 

As the photogeneration of the sample with additive was voltage independent, as shown in Fig.~\ref{fig:tdcf}, the respective generation current $j_{gen}$ was assumed to be constant and set equal to the short circuit current density,
 \begin{equation}
j_{gen}\approx j_{sc},
\label{eq:jgen}
\end{equation}
similar to the approach in Ref.~\onlinecite{shuttle2010}, \onlinecite{gluecker2012}.

\begin{figure}[tb]
	\includegraphics[width=0.9\linewidth]{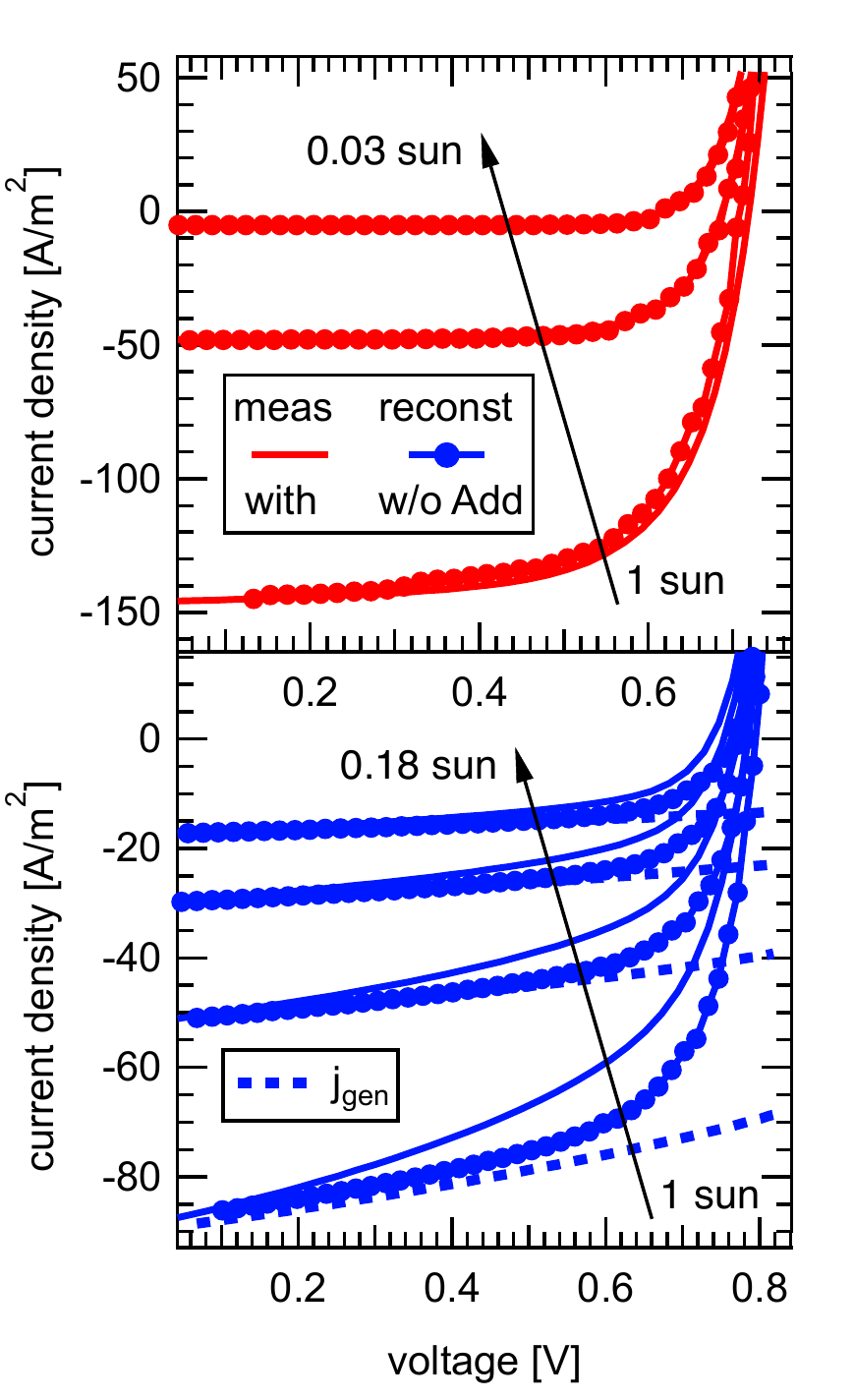}
	\caption{Measured and reconstructed current/voltage response of PTB7:PC$_{71}$BM organic solar cells processed with (top) and without (bottom) the additive DIO. The deviation between direct measurement and reconstruction for the latter is discussed in the text.}
	\label{fig:reconstr}
\end{figure}

Instead, for the solar cell fabricated from pure CB solution, the voltage dependent polaron pair dissociation PP($V$) derived by TDCF is substantial. It was considered for the reconstruction by a voltage dependent generation current 
\begin{equation}
	j_{gen} (V)=j_{sc}\cdot \text{PP}(V),
	\label{eq:jgenV}
\end{equation}
with the term PP($V$) derived from the polynomial fit (Fig.~\ref{fig:tdcf} dashed line), accounting for the relative charge photogeneration between $V=0$~V and $V=V_{oc}$. In order to calculate the $j/V$ response for both device types, either Eq.~(\ref{eq:jgen}) or Eq.~(\ref{eq:jgenV}) were entered into Eq.~(\ref{eq:j}) for the device processed with or without additive, respectively. Thus gained $j/V$ data perfectly agree with the measured data for the device with additive for three light intensities presented (Fig.~\ref{fig:reconstr}~top). Instead, although experimentally determined geminate and nongeminate losses were considered, the reconstruction of the device processed from pure CB shows significant deviations from the measured $j/V$ characteristics (Fig.~\ref{fig:reconstr}~bottom).

\section{Discussion}

The application of the co-solvent DIO results in a dramatic change of the active layer morphology~\cite{collins2012, kraus2012} and also in a considerably improved device performance (Fig.~\ref{fig:iv}). For the present material systems the additive leads to a more uniform morphology and hierarchical nanomorphologies and the formation of an interpenetrating network in accordance with earlier reports~\cite{liang2010b, chen2011}. For a device processed without DIO a large phase separation is present and isolated fullerene domains can be assumed from AFM/TEM images, instead~\cite{liang2010b, lou2011, collins2012}. Despite these differences, our TPV and CE studies revealed similar charge carrier lifetimes and, thus, comparable nongeminate recombination rates for both devices at low charge carrier densities. This is consistent with the results from grazing incidence X-ray scattering and resonant soft-X-ray scattering studies, which showed that the addition of DIO to the casting solution has little effect on the overall domain crystallinity or on the domain composition~\cite{chen2011, collins2012}. Nevertheless, for carrier concentrations above $10^{22}$m$^{-3}$ the lifetimes differ. In order to determine the correct nongeminate recombination yield by TPV/CE measurements, it was assumed that all charges participating in the recombination process are encompassed by the CE procedure. This assumption can be critical if isolated fullerene domains are present, which is discussed below.

Concerning the charge photogeneration, from prebias dependent TDCF measurements a field-assisted dissociation of polarons is found for the device processed without additive, while the device prepared with DIO shows a rather constant separation yield (Fig.~\ref{fig:tdcf}). The field dependence is rather weak even for the former device: we find about 12\% less generation close to $V_{oc}$ than under short circuit conditions. 

Field independent generation in polymer--fullerene blends was attributed to the co-existence of mixed and pure domains, whereby an energy gradient is established that drives the photogenerated charges out of the intermixed regions~\cite{jamieson2012, shoaee2013}. Recent TDCF experiments on another polymer:fullerene blend revealed that larger and purer domains lead to more efficient and less field-dependent photogeneration~\cite{albrecht2012b}. In PTB7:PC$_{71}$BM blends processed without the additive, the fullerene domains were shown to be very pure. Processing with the additive reduced mainly the domain size but did not affect its composition~\cite{collins2012}. The rather weak or even absent field dependence of generation seen here is consistent with this structural picture. Moreover, inefficient exciton harvesting in combination with field-assisted free carrier formation within the individual large domains, as has been recently suggested by Burkhard et al.\cite{burkhard2012}, might account for the weaker and slightly bias-dependent $Q_{tot}(V)$ in the blend processed without DIO. 

For the device with additive, reconstructed and measured $j/V$ data coincide almost perfectly, as shown in  Fig.~\ref{fig:reconstr}~top: Both, fill factor and $V_{oc}$ are reproduced quite accurately for all light intensities. As only nongeminate losses were considered to calculate the $j/V$ behavior, they are identified as the dominant loss process responsible for device performance limitation. 

In case of the device processed from pure CB, both field dependent photogeneration and nongeminate recombination limit the performance. However, despite considering both loss mechanisms, the reconstructed $j/V$ data overestimates the device performance (Fig.~\ref{fig:reconstr}~bottom).
This discrepancy stems from an underestimation of either geminate or nongeminate losses. The degree of geminate recombination was verified by TDCF measurements with an excitation wavelength of 500~nm at different delay times (10~ns up to 50~ns) and excitation intensities, but the determined PP($V$) dependence remained virtually unchanged. Thus, we expect PP($V$) and, therefore, the photogeneration current $j_{gen}$($V$) to be determined correctly, at least as long as the dissociation of polaron pairs takes place within 50 ns. It is known from other low bandgap polymers like PCPDTBT or PCDTBT that CT decay times of less than 10~ns is a credible assumption\cite{Jarzab2011, etzold2011}. 

In order to discuss the reliability of the reconstruction, we point out the assumptions and related uncertainties. For determining the nongeminate loss current $j_{loss}$, we use the experimentally determined nongeminate recombination rate $R(n)$ from TPV/CE measurements under open circuit conditions. It is assumed not to be explicitly dependent on voltage. A voltage dependent recombination rate might be due to a field dependent mobility, with a negative voltage coefficient in some blend compositions~\cite{koster2010, albrecht2012}. Moreover, spatial variations of the charge carrier density cannot be resolved in CE experiments. If gradients are present, they are more pronounced towards $V=0$~V than around $V_{oc}$~\cite{deibel2009b}. However, both effects---a potentially field dependent mobility and significant charge carrier gradients---would lead to an overestimation of nongeminate losses, not an underestimation. Therefore, they cannot be responsible for the apparent deviation of device performance. 

An additional factor potentially influencing the reconstruction is the wavelength dependence of photogeneration. The field-dependence of the photocurrent was studied by Brenner \etal based on external quantum efficiency measurements with bias light, i.e., including contributions of geminate and nongeminate recombination without distinguishing between them. They observed a slightly stronger field-dependence of photocurrent generation in PTB7 relative to the fullerene for devices processed without additive~\cite{brenner2011}. A comparison of the voltage regime relevant for our reconstruction, from 0~V to positive bias, shows that the maximum relative deviation between the photocurrent at 500~nm (dominant PC$_{71}$BM absorption) and 650~nm is less than 5\%. We could verify these results by own bias-dependent EQE measurements and included them in the supporting information (SI).

A relevant scenario based on our results and the reports of morphology in literature~\cite{liang2010b, lou2011, collins2012} briefly discussed above, however, could explain the discrepancy between $j/V$ characteristics and reconstruction for the device without additive. It is based on the presence of trapped charges (photogenerated or intrinsic) in spatially isolated domains, i.e., without percolation pathways to the respective electrode. These trapped charge carriers could recombine with mobile ones at the organic--organic interfaces influencing the apparent recombination rate. However, they could not contribute directly to the charge transport, and could not therefore be observed in charge extraction experiments. Consequently, the calculated nongeminate loss current $j_{loss}$ (see Eq.~(\ref{eq:nong}, \ref{eq:j})) would be underestimated, leading to an overestimated reconstruction of the photocurrent. We believe this scenario to be likely for PTB7:PC$_{71}$BM solar cells prepared without the co-solvent. We point out that recently for two other bulk heterojunction systems, the role of isolated domains without percolation pathways and the resulting negative impact on the solar cell performance was discussed.\cite{bartelt2013,ma2013} 

Indeed, a combination of resonant X-ray scattering and microscopy as well as AFM images revealed 50-200~nm pure PC$_{71}$BM domains~\cite{collins2012, kraus2012} favoring a trapping process as described above. In Fig.~\ref{fig:n_V}, theoretical charge carrier density data $n(V)$ required for a successful $j/V$ reconstruction of 1 sun illumination was exemplarily added for the device processed without additive (blue dashed line). Also, the correspondingly shifted effective lifetime in dependence on the reconstructed charge carrier density at $V_{oc}$, $\tau(n)$, is shown in Fig.~\ref{fig:tau_n} for the same device (blue open circles). Both representations illustrate that a considerable amount of trapped charges contributes to the nongeminate loss current. Further evidence that trapped charges are the origin of the apparent deviation is given in the SI. There, it is shown that the determined dark capacitance $C_{dark}$ is almost 50\% higher than the estimated geometric capacitance $C_{geo}$ of a dielectric with $\epsilon=3.7$. 
As described in the SI, C$_{geo}$ is required to correct for charges on the electrodes, and leads to charge carrier densities in the same range as necessary for successful $V_{oc}$ reconstruction. Therefore, the difference between $C_{dark}$ and $C_{geo}$ is attributed to charge carriers which are spatially trapped and cannot be extracted during the charge extraction experiment, supporting the scenario with trapped charge carriers on isolated fullerene domains.


\section{conclusion}
PTB7:PC$_{71}$BM bulk heterojunction solar cells prepared from pure CB solution and from a CB/DIO mixture were analyzed
by voltage dependent CE, TPV and TDCF measurements to elucidate the origin of performance limitation. In devices processed with DIO, a voltage independent charge photogeneration was found. We performed a $j/V$ reconstruction, and found that it agrees very well with the measured response. This finding allows us to identify nongeminate recombination as the performance limiting loss mechanism for the highly efficient decice with DIO. In contrast, devices processed from pure CB solution and yielding lower efficiency show both, severe geminate and nongeminate losses. There we found a strong deviation of measured and reconstructed $j/V$ characteristics, which we discussed with respect to spatially trapped charge carriers in isolated PC$_{71}$BM domains. We show that these trapped charge carriers can explain the discrepancy in the $j/V$ reconstruction, and support this interpretation by measurements of the dark capacitance and comparison with the geometric capacitance.

\section{Experimental Section}	\label{exp}

Bulk heterojunction (BHJ) solar cells based on PTB7:PC$_{71}$BM were prepared by spin coating a 35~nm thick layer of poly(3,4-ethylendioxythiophene):polystyrolsulfonate (Clevios P VP AI 4083) on indium tin oxide samples with post-annealing step of 130$^{\circ}$C~for 10 minutes in a water-free environment. The PTB7:PC$_{71}$BM 1:1.5 blend is spin coated in inert atmosphere from a chlorobenzene (CB) solutions of 20~mg/ml to realize a 125~nm thick layer. In case of the device denoted as with additive 3vol\% 1,8-diiodooctane (DIO) were used as co-solvent in the CB solution. Finally, the top metal contacts (Ca/Al) were evaporated thermally on top of the organic layer defining the active area of $A=9.2$~mm$^2$. PTB7 was purchased from 1-material and PC$_{71}$BM was supplied from Solenne. All materials were used without further purification.

Prior to any additional measurements an Oriel 1160 AM1.5G solar simulator was used to perform illuminated $j/V$--measurements of devices kept under inert glove-box atmosphere. Further static and transient electrical studies were carried out in a closed cycle optical cryostat. For the TPV measurements the organic solar cells were connected to a digital storage oscilloscope (Agilent Infiniium DSO90254A) via a 1.5~G$\Omega$ input resistance of a voltage amplifier. A high power white light emitting diode (Cree) was used to illuminate the active area of the devices, causing a constant rate of generation and recombination. An Nd:YAG laser pulse (532~nm excitation wavelength, pulse duration 80~ps) is applied to generate an additional amount of charges causing a small perturbation of the photovoltage. The photogenerated charge was determined from charge extraction transients. The externally applied voltage is supplied to the solar cell by a Keithley 2602 in combination with a fast digital/analog switch. The derived experimental data were treated as reported in Ref.~\onlinecite{foertig2012b}, \onlinecite{rauh2011}, while all charges were corrected for recombination losses during extraction as mentioned above. 

TDCF experiments were performed by illuminating the solar cell with a short Nd:YAG laser pulse (NT242 EKSPLA, 5.5~ns pulse width, 500~Hz repition rate, 500~nm excitation wavelength) at different applied pre-biases as described in Ref.~\onlinecite{albrecht2012}. The photogenerated charge carriers were extracted by applying a high rectangular voltage pulse with a pulse generator (Agilent 81150A) in reverse direction. The current through the device was measured with a Yokogawa DL9140 oscilloscope via a 50~$\Omega$ input resistor. The pixel size of the active area was 1~mm$^2$ to avoid RC-time limitations. 
The total photogenerated charge $Q_{tot}$ is determined by the integral of the current transient.

\begin{acknowledgments}
The authors thank Andreas Zusan for helpful discussions and reading the manuscript and Ilja Lange for cell preparation. This work was supported by the Deutsche Forschungsgemeinschaft in the framework of the project DE 830/11-1 within SPP~1355. C.D. gratefully acknowledges the support of the Bavarian Academy of Sciences and Humanities. V.D.s work at the ZAE Bayern is financed by the Bavarian Ministry of Economic Affairs, Infrastructure, Transport and Technology.
\end{acknowledgments}

\end{document}